\documentstyle[sprocl,epsfig]{article}

\def\be{\begin{equation}}
\def\ee{\end{equation}}
\def\bea{\begin{eqnarray}}
\def\eea{\end{eqnarray}}
\def\gsim{ \lower .75ex \hbox{$\sim$} \llap{\raise .27ex \hbox{$>$}} }  
\def\lsim{ \lower .75ex\hbox{$\sim$} \llap{\raise .27ex \hbox{$<$}} }  

\begin{document}

\title{INTERNAL SHOCKS AND THE BLAZAR SEQUENCE}

\author{Dafne Guetta}
\address{Osservatorio Astrofisico di Arcetri, Largo E. Fermi 5, I--50125 Firenze, Italy}

\author{Gabriele Ghisellini}
\address{Osservatorio Astronomico di Brera, via Bianchi 46, I--23807 Merate, Italy}

\author{Davide Lazzati}
\address{IoA, University of Cambridge, Madingley Road, Cambridge CB3 0HA, UK}

\author{Annalisa Celotti}
\address{SISSA/ISAS, via Beirut 2-4, I--34014 Trieste, Italy}


\maketitle\abstracts{
We consider the internal shock model as the dissipation mechanism
responsible for the emission in blazars. It has been shown that this
model is successful in reproducing the observed spectral energy
distribution and the variability properties of a powerful blazar like
3C 279. However, the blazar family covers a wide range of spectral
characteristics which appear to be correlated and the whole class can
be seen as a sequence: the frequency and the intensity of the low
energy versus high energy peak intensity increase with decreasing
luminosity. We show that the internal shock model can satisfactorily
account also for the properties of the low power blazars like BL Lac
and Mkn 421 and it is successful in reproducing the blazar sequence.}

\section{Introduction}

The discovery that blazars are strong $\gamma$-ray emitters together
with the results from the multiwavelength campaigns of these sources
have allowed to deepen our knowledge on these objects. We know that
the Spectral Energy Distribution (SED) of blazars is characterized by
two broad emission peaks \cite{F98} strongly variable on different
timescales \cite{WW95}. Two main radiation processes produce these
peaks, i.e. the synchrotron at low frequencies and the inverse Compton
at high energies (see e.g. \cite{S94} for a review).  The
relative importance of the two peaks and their location in frequency
appear to be a function of the total power of blazars \cite{F98} 
\cite{G98} leading to a {\it blazar sequence}. The blazar emission is
variable on different timescales, which are typically weekes-months
for the radio emission of the order of a day for the $\gamma$-ray. 

Several studies, mainly based on the fitting of the observed SED, have
allowed us to consistently derive the physical parameters in the
emitting region of these sources, however many aspects in the
understanding of relativistic jets remain open, most notably the jet
energetics and particle acceleration.  In order to explore these
issues and their relationship we have considered a scenario where the
plasma characteristics are not treated as free parameters, but are the
result of the jet dynamics, thus relating the observed emission
properties to the transport of energy along the jet.  We have achieved
that by {\it quantitatively} considering a dissipation mechanism
within the jet responsible for the emission of blazars analogous to
the standard mechanism proposed to explain gamma--ray bursts
\cite{RM94}: the internal shock model. This model was actually
originally proposed for blazars more than two decades ago \cite{R78}.
The most important assumption of the model is that the central power
engine produces energy which is channeled into jets in an intermittent
way, though such a time dependent process cannot be easily inferred
from first principles. The model assumes that different parts of the
jet (shells), moving at different speeds can collide producing shocks,
giving rise to the non-thermal radiation we see. This mechanism has a
limited efficiency because only a small fraction of the bulk energy
can be converted into radiation (unless the contrast between the two
initial Lorents factor is huge, see Beloborodov 2000 \cite{B00}; Guetta
et al. 2001 \cite{G01}; but also Ghisellini 2002 \cite{G02}).  
In blazars, on the other hand, the radiative output has to be a small
fraction of the energy transported by the jet (less than 10\%), since
most of the bulk energy has to go un--dissipated to the outer
radio--lobes.  The low efficiency required for blazars is a
characteristic of the model, moreover the internal shock model can
naturally explain other properties of these sources: a minimum
dimension for the $\gamma$--ray source is required, as well as a minimum distance 
from the accretion disk, in order to avoid pair production.
The internal shock model can naturally account for
this, since the jet becomes radiative at ($\gsim 10^{16}- 10^{17}$ cm)
from the black hole.
The observed large amplitude variability must be explained by a non steady
state model: we believe that the shell--shell collision scenario is the
simplest among them.  The shells that have already collided once can
collide again, at larger distances from the black hole with reduced
efficiency, since the Lorentz factor contrast of the colliding shells
decreases. This can explain why the luminosity of jets decrease with
distance. Fig. 1 shows the efficiency as a function of distance for a
simulation we have done for Mkn 421 \cite{GG02}.
A detail study via numerical simulations of the predictions of this
model in blazars has been carried out by \cite{S01} 
for powerful objects (like 3C279). However the SED of this
object does not represent the whole of the blazar family, which covers
a wide range of spectral characteristics (luminosity, frequency of the
peaks of emission and their relative intensity). We have applied our
model to Mkn 421 and BL Lac itself which, besides being among the best
studied BL Lac objects, are representative of the extremely blue and 
intermediate blazar class, respectively. Our aim is to see if the
proposed scenario can explain the difference of the SED along the
blazar sequence, more than to explain every detail of a particular
spectrum of a specific object.
\begin{figure}
\epsfysize=13cm 
\epsfbox{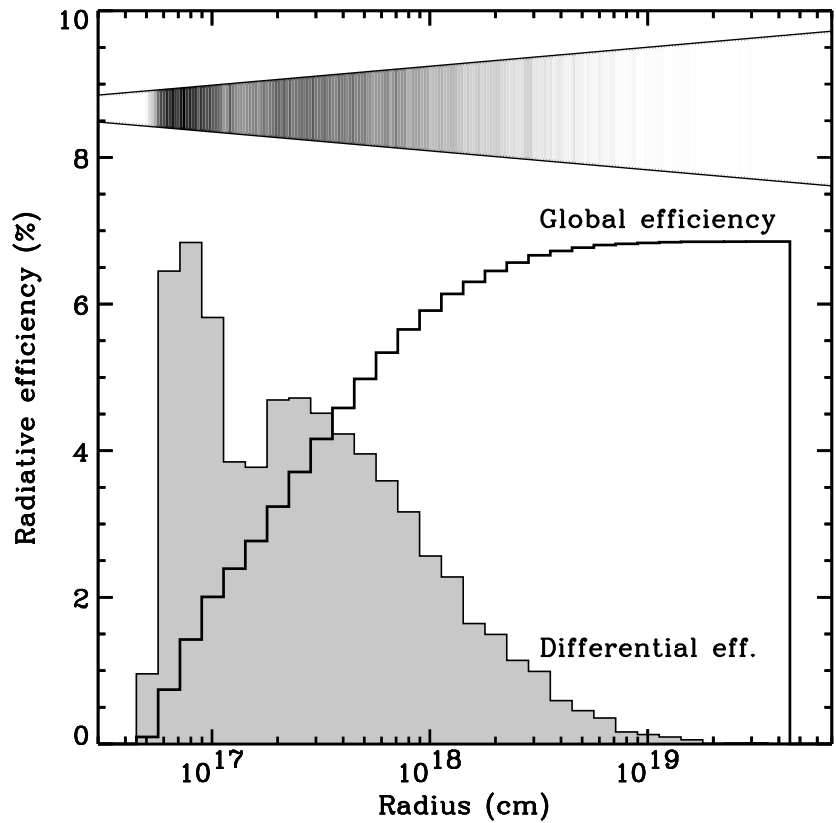}
\vskip -1 true cm
\caption{
The result of a simulation of the internal shock model for Mkn
421. The average local radiative efficiency and the cumulative
efficiency versus the distance from the black hole. The first ``peak''
in the efficiency corresponds to the first collisions between shells,
while the ``tail'' at larger distances corresponds to second, third
(and so on) collisions. The top of the plot shows a schematic
representation of the jet, with grey levels proportional to the local
efficiency. From Guetta et al. (2002), in prep.}
\end{figure}

\section{Internal shocks and spectrum}
\subsection{Generalities}

In this work we use the approximate model of the unsteady wind
described in detail in Spada et al. (2001) and in the following we
report only its most relevant assumptions. We consider a compact
source, of typical dimension $R_0\sim 10^{14}$ cm, which produces an
unstable relativistic wind characterized by an average luminosity
$L_{\rm jet}$. 
The emission from the jet is obtained by adding pulses
radiated in a series of internal shocks that occur in the wind.
The wind flow is described as a sequence of $N=t_w/t_v$ shells, where
$t_w$ is the overall duration of the emission of the wind and 
$t_v\ll t_w$ is the average interval between two consecutive shells. 
Each shell is characterized by four parameters
(the subscript $j$ denotes the $j_{\rm th}$ shell):
the ejection time $t_j$,
randomly extracted from a uniform distribution with an average value of $t_v$;
the Lorentz factor $\Gamma_j$ that is randomly extracted from a distribution
between $\Gamma_m$ and $\Gamma_M$; 
the mass $M_j$ also randomly extracted;
the width $\Delta_{j}\sim R_0$.
The dynamics of the wind
expansion is characterized by a series of two--shell collisions in
which the faster shells catch up with the slower ones in the outer
parts of the ejecta.  The energy released in each shock is assumed to
be distributed among electrons and magnetic field with fractions
$\epsilon_e$ and $\epsilon_B$, respectively.  In this way we can
determine the magnetic field strength needed to evaluate the
synchrotron and the inverse Compton emission.  Electrons are assumed
to be accelerated at relativistic energies with a power law spectrum.
\subsection{Local spectra} 
We refer the reader to Spada et al. (2001) for a detailed description
of the derivation of the spectra that is obtained considering the
synchrotron, the synchrotron self-Compton (SSC) and the external
Compton radiation mechanisms.  The only thing we would like to point
out is that even in the case of low power BL Lacs we consider the
presence of a source of soft photons external to the jet, which we
identified with the emission reprocessed in the broad line region
(BLR).  We consider a luminosity $L_{\rm ext}=a L_{\rm disk}$ (with $a\sim 0.1$) 
produced
within $R_{\rm BLR}$, corresponding to a radiation energy density (as
measured in the frame comoving with the shell) $U_{\rm ext} = (17/12)
aL_{\rm disk} \Gamma^2 /(4 \pi R_{\rm BLR}^2 c)$.  (e.g. \cite{GM96}).  For simplicity, this seed photon component
is considered to abruptly vanish beyond $R_{\rm BLR}$.  High and low
power blazars have different line luminosities (and different
photoionizing disk continua).  According to Kaspi et al. (2000)
\cite{K00}, $R_{\rm BLR}$ and $L_{\rm disk}$ are related by $R_{\rm
BLR}\propto L_{\rm disk}^b $ with $b\sim 0.7$ This implies that
blazars with weaker broad emission lines should have smaller broad
line regions.  In turn, this implies that, in low power BL Lacs, the
first collisions between shells can occur preferentially outside
$R_{\rm BLR}$.

\section{Results and Discussion}


We applied our model to Mkn 421 and BL Lac itself. 
In Tab. 1 we list the input parameters used
for the model, and for comparison we also report the
same parameters for 3C 279. Via the numerical simulations the full
time--resolved spectral behavior of the sources can be determined.
The full time--dependent behavior of the simulated sources can be also
examined through animations 
(see {\it http://ares.merate.mi.astro.it/$\sim$gabriele/421/index.html}) 
in which the temporal and spectral evolutions are simultaneously shown.

The results of this work show that the internal shock model can
satisfactorely account also for the properties of the low power
blazars.  
We find that the key model parameters that need to be changed
in order to reconstruct the sequence in the frame of the internal
shock scenario, are indeed the broad lines intensity and the power
carried in the jet which is proportional to the radiative luminosity.
These parameters in turn regulate the SED shape, as they control the
cooling efficiency of the emitting particles.  The internal shock
scenario in fact determines a characteristic time interval for the
injection of relativistic electrons during each shell--shell
collision, of the order of the dynamical timescale $t_{cross}$.  
The intensity of the spectrum out of each collision is maximized at the
end of this particular timescale.  
Two regimes are relevant: {\it fast} and {\it slow} cooling, corresponding 
to whether electrons of energy $\gamma_b$ ($\gamma_b m_e c^2$
is the minimum energy of the injected electrons)
can (fast cooling) or not (slow cooling)
radiatively cool in $t_{cross}$. 
In highly powerful blazars the fast
cooling regime applies in the inner regions (within the BLR) where
also most of the power is dissipated. Consequently the peak
frequencies are produced by the electrons of energy 
$\gamma_{b} m_{e} c^2$ (as even these electrons can cool in $t_{cross}$).  
In the lowest
powerful blazars instead the slow cooling regime applies to {\it all}
collisions. Consequently only the electrons with the highest energy
can cool in $t_{cross}$ (even for the most powerful collisions): the
peak frequencies thus shift to very high values. Between these two
extremes there is the possibility of BL Lac objects with broad lines
of intermediate intensity originating at a distance from the accretion
disc within which very few (but not zero) shell-shell collisions can
occasionally take place.  This is the case of BL Lac itself and all
the blazars of intermediate luminosity.  Observationally this
corresponds to a SED produced by internal shocks of moderate Compton
to synchrotron luminosity ratio as the seed photons for Compton
scattering are provided only by the synchrotron photons (collisions
outside the BLR).  The rare collisions within the BLR will give rise
to a dramatic change in the SED characterized by a large increase of
the Compton component as in the case of BL Lac itself. \\ One can
conclude that the internal shock scenario explains the blazar sequence
and the main characteristics of the blazar behaviour. While
observationally we were able to see that the important quantity is the
ratio between the observed luminosity in the jet and the disc (thus
BLR), the internal shock scenario allows to i) directly connect the
radiated jet luminosity with the effective power carried by the jet
and ii) account for the preferred distance at which most of the jet
luminosity is produced. While the latter distance is qualitatively the
same for high and low power blazars, the BLR is instead located at
very different distances in the two sub--classes of objects, as
determined by the ionizing luminosity.
\begin{table} 
\begin{center} 
\begin{tabular}{lllllllllll} 
\hline 
Source   
& $\langle L_{\rm jet}\rangle$   
& $\Gamma_{\rm m} $--$\Gamma_{\rm M}$ 
& $ t_{\rm v}$ 
& $ \epsilon_{\rm e}$ 
& $ \epsilon_{\rm B}$ 
& $ L_{\rm BLR}$ 
& $ R_{\rm BLR}$\\  
 &erg s$^{-1}$ & & s & & &erg s$^{-1}$ & cm  \\ 
\hline 
Mkn 421 &8e45 &10--25 &1.3e4  &0.5 &6e-4 &1e42  &1e16  \\ 
BL Lac  &4e46 &10--25  &1.6e4  &0.5 &1e-2 &8e42  &7e16  \\ 
3C 279  &1e48 &10--25 &1.0e4  &0.5 &4e-3 &1e45  &5e17  \\  
\hline 
\end{tabular} 
\caption{Input parameters of the model for the 3 blazars. 
} 
\end{center} 
\label{tab:fits} 
\end{table}

\end{document}